\documentclass[preprint,preprintnumbers,amsmath,amssymb]{revtex4}

\usepackage{bm}

\begin{document}

\preprint{APCTP 2005-10}
\preprint{hep-ph/0512294}

\title{Neutrino mixing and CP violation from\\  
    Dirac-Majorana bimaximal mixture and quark-lepton unification}

\author{Junpei Harada}
 \affiliation{Asia Pacific Center for Theoretical Physics, Pohang 790-784, Korea}
 \email{jharada@apctp.org}


\begin{abstract}
We demonstrate that only two ansatz can produce the features of the neutrino mixing angles. The first ansatz comes from the quark-lepton grand unification; $\nu_{Di} = V_{CKM} \nu_{\alpha}$ is satisfied for left-handed neutrinos, where $\nu_{Di}\equiv (\nu_{D1},\nu_{D2},\nu_{D3})$ are the Dirac mass eigenstates and $\nu_{\alpha}\equiv (\nu_e, \nu_\mu, \nu_\tau)$ are the flavour eigenstates. The second ansatz comes from the assumption; $\nu_{Di} = U_{bimaximal} \nu_{i}$ is satisfied between the Dirac mass eigenstates $\nu_{Di}$ and the light Majorana neutrino mass eigenstates $\nu_{i}\equiv (\nu_1, \nu_2, \nu_3)$, where $U_{bimaximal}$ is the $3 \times 3$ rotation matrix that contains two maximal mixing angles and a zero mixing. By these two ansatz, the Maki-Nakagawa-Sakata lepton flavour mixing matrix is given by $U_{MNS} = V_{CKM}^\dagger U_{bimaximal}$. We find that in this model the novel relation $\theta_{sol} + \theta_{13} = \pi/4$ is satisfied, where $\theta_{sol}$ and $\theta_{13}$ are solar and CHOOZ angle respectively. This "Solar-CHOOZ Complementarity" relation indicates that only if the CHOOZ angle $\theta_{13}$ is sizable, the solar angle $\theta_{sol}$ can deviate from the maximal mixing. Our predictions are $\theta_{sol} = 36^\circ$, $\theta_{13} = 9^\circ$ and $\theta_{atm} = 45^\circ$, which are consistent with experiments. We also infer the CP violation in neutrino oscillations. The leptonic Dirac CP phase $\delta_{MNS}$ is predicted as $\sin \delta_{MNS} \simeq A \lambda^2 \eta$, where $A, \lambda, \eta$ are the CKM parameters in Wolfenstein parametrization. In contrast to the quark CP phase $\delta_{CKM} \simeq {\cal O}(1)$, the leptonic Dirac CP phase is very small, $\delta_{MNS} \simeq 0.8^\circ$. Furthermore, we remark that the ratio of the Jarlskog CP violation factor for quarks and leptons is important, because the large uncertainty on $\eta$ is cancelled out in the ratio, $R_J \equiv J_{CKM}/J_{MNS} \simeq 4\sqrt{2} A \lambda^3 \simeq 5 \times 10^{-2}$. 
\end{abstract}

\maketitle

In the past several years, our knowledge about the lepton flavour structure has been drastically improved by the progress of neutrino oscillation experiments~\cite{Fukuda:1998tw,Fukuda:1998fd,Ahmad:2001an,Eguchi:2002dm,Apollonio:1997xe,Cleveland:1998nv,Hampel:1998xg,Ahn:2001cq}. Theoretical lepton flavour models have been extensively studied (for a recent review, see e.g.,~\cite{Altarelli:2004za} and references therein), however, there is still no standard answer.
Therefore, trying to understand the lepton flavour structure is a very important theoretical challenge in particle physics today.

In this Letter we demonstrate that only two ansatz can produce the neutrino mixings. Two ansatz will be discussed soon. We find a novel relation $\theta_{sol} + \theta_{13} = \pi/4$ between the solar mixing angle $\theta_{sol}$ and the CHOOZ angle $\theta_{13}$. We call this relation the "Solar-CHOOZ Complementarity" in this Letter. We also infer the CP violation in neutrino oscillations. We remark that the ratio of the Jarlskog CP $J$ factor for quarks and leptons is important, because it does not depend on the CP violation parameter with large uncertainty.

We write the lepton flavour mixing matrix
\begin{eqnarray}
 \nu_{\alpha} = U_{\alpha i} \nu_i,   \label{eq:MNS}
\end{eqnarray}
where $\nu_\alpha \equiv (\nu_e,\nu_\mu,\nu_\tau)$ are the flavour eigenstates and $\nu_i \equiv (\nu_1, \nu_2, \nu_3)$ are the light Majorana mass eigenstates. The $3 \times 3$ unitary matrix $U_{\alpha i}$ is the Maki-Nakagawa-Sakata (MNS) lepton flavour mixing matrix $U_{MNS}$~\cite{Maki:1962mu}. The current status of the neutrino oscillation parameters is elegantly summarized in~\cite{Maltoni:2004ei}.

The first ansatz in our approach comes from the quark-lepton grand unification~\cite{Pati:1974yy}. We write the first ansatz
\begin{eqnarray}
 \nu_{D i} = V_{CKM} \nu_\alpha, \label{eq:1ansatz}
\end{eqnarray}
where $\nu_{Di} \equiv (\nu_{D1},\nu_{D2},\nu_{D3})$ are Dirac mass eigenstates and $\nu_\alpha$ are the flavour eigenstates. The reason that we expect this ansatz is quite simple. For the quark sector, in the Standard Model $u = V_{CKM} u^\prime$ is satisfied for the left-handed quarks, where $u \equiv (u,c,t)$ are the Dirac mass eigenstates and $u^\prime \equiv (u^\prime, c^\prime, t^\prime)$ are the flavour eigenstates. The $V_{CKM}$ is the Cabibbo-Kobayashi-Maskawa (CKM) quark flavour mixing matrix~\cite{Cabibbo:1963yz}. In the quark-lepton grand unification, the Dirac mass structure can be same for quarks and leptons. Therefore, the same Dirac flavour relation would be satisfied between quarks and leptons, and the first ansatz~(\ref{eq:1ansatz}) can be naturally expected. Because of this ansatz, the neutrino mixing angles and the leptonic Dirac CP phase can be given in terms of the CKM quark mixing parameters.

Before discussing the second ansatz, we give some comments on the first ansatz. If you are experimentalists, you can skip this paragraph and go to the second ansatz. We know that the up quark mass matrix equals to the Dirac neutrino mass matrix in SO(10) or $E_6$ grand unification. And the down quark mass matrix relates to the charged lepton mass matrix in SU(5), SO(10) or $E_6$ grand unification. Therefore, in SO(10) or $E_6$ grand unification, the first ansatz~(\ref{eq:1ansatz}) can be naturally expected. However, in these grand unified models, it is well known that the CKM mixing matrix is predicted as $V_{CKM} = {\bf 1}$, because all the standard model fermions (+ right-handed neutrinos) are unified into the same gauge multiplets. We can overcome this difficulty, and construct the realistic models that predict $V_{CKM} = {\bf 1} + \Delta$ by 1) adding the higher dimensional Higgs multiplets~\cite{Mohapatra:1986uf} and/or 2) considering the nonrenormalizable higher dimensional interactions~\cite{Albright:1998vf} and/or 3) using the twisting mechanism in $E_6$ unified models~\cite{Bando:1999km}.  Such explicit model buildings are important studies, however, in this Letter we take a somewhat different view. Instead of specifying scenarios of the unified models, we only impose the first ansatz~(\ref{eq:1ansatz}) that is motivated by the quark-lepton grand unification.

The second ansatz comes from the assumption. We write the second ansatz
\begin{eqnarray}
 \nu_{D i} = U_{bimaximal} \nu_{i},  \label{eq:2ansatz}
\end{eqnarray}
where $\nu_{Di}$ are the Dirac mass eigenstates and $\nu_i$ are the light Majorana mass eigenstates. The $3 \times 3$ rotation matrix $U_{bimaximal}$ contains two maximal mixing angles (1-2 mixing and 2-3 mixing) and a zero 1-3 mixing. This second ansatz means that the dimension 5 effective interaction $y_{ij}L_i L_j H H/M$ is diagonalized by $U_{bimaximal}$, where $L_{i}$ is the SU(2) lepton doublet in the basis of $\nu = \nu_{D i}$ and $H$ is the Higgs doublet. $y_{ij}$ are Yukawa couplings and $M$ is an energy scale to be understood as the right-handed Majorana neutrino mass scale because of the neutrino seesaw mechanism~\cite{Seesaw}.

Before further discussions, we give the reason why we impose the second ansatz by assumption. First of all, even in any grand unified models (for example, SU(5), SO(10) or $E_6$), in general there is no theoretical relation between the Dirac mass eigenstates $\nu_{Di}$ and the light Majorana mass eigenstates $\nu_i$, if we do not impose hypothetical flavour symmetries or texture. Therefore, in this Letter we impose the simplest assumption~(\ref{eq:2ansatz}) that is motivated by the neutrino oscillation experiments~\cite{Fukuda:1998tw,Fukuda:1998fd,Ahmad:2001an,Eguchi:2002dm,Apollonio:1997xe,Cleveland:1998nv,Hampel:1998xg,Ahn:2001cq}, and demonstrate that the assumption~(\ref{eq:2ansatz}) and the first ansatz~(\ref{eq:1ansatz}) are sufficient to produce the neutrino mixings. Theoretical models with flavour symmetries that produce the bimaximal mixing matrix have been studied~\cite{Petcov:1982ya}, however, in this Letter we do not discuss such symmetries.

From Eq.~(\ref{eq:MNS}) and two ansatz~(\ref{eq:1ansatz}) and (\ref{eq:2ansatz}), we can write the neutrino mixing matrix
\begin{eqnarray}
 U_{MNS} = V_{CKM}^\dagger U_{bimaximal}. \label{eq:MNS2}
\end{eqnarray}
This equation shows how neutrino mixing angles and the leptonic Dirac CP phase can be given in terms of the CKM quark mixing parameters. We do not consider the Majorana CP phases in this Letter, which do not affect neutrino oscillations. In our approach, Eq.~(\ref{eq:MNS2}) comes from only two ansatz~(\ref{eq:1ansatz}) and~(\ref{eq:2ansatz}). We comment that the similar relation was studied from different motivations and by different ways~\cite{Giunti:2002ye,Rodejohann:2003sc,Datta:2003qg,Xing:2005ur,Dutta:2005bb}.

In this Letter we consider the CKM mixing matrix in Wolfenstein parametrization~\cite{Wolfenstein:1983yz}. From Eq.~(\ref{eq:MNS2}), the neutrino mixing angles can be expressed in very simple form
\begin{eqnarray}
 \sin^2 2\theta_{atm}  = 1 + {\cal O}(\lambda^4), \quad
 \sin^2 2\theta_{sol} = 1 - 2\lambda^2  + {\cal O}(\lambda^4), \quad
 \sin^2 \theta_{13} = \frac{\lambda^2}{2} + {\cal O}(\lambda^4),  \label{eq:angles}
\end{eqnarray}
where $\lambda \equiv \sin \theta_C$. We find that Eq.~(\ref{eq:angles}) leads to a novel relation between the solar angle $\theta_{sol}$ and the CHOOZ angle $\theta_{13}$
\begin{eqnarray}
 \theta_{sol} + \theta_{13} = \frac{\pi}{4} + {\cal O}(\lambda^3), \label{eq:sol-13}
\end{eqnarray}
where $\lambda^3$ radians $\simeq 0.6^\circ$ for $\lambda = 0.22$~\cite{Eidelman:2004wy}.
This relation is similar to the Quark-Lepton Complementarity~\cite{Minakata:2004xt}, which refers the relation $\theta_{sol} + \theta_C = \pi/4$ between the solar angle $\theta_{sol}$ and the Cabibbo angle $\theta_C (\simeq 13^\circ$~\cite{Eidelman:2004wy}). In this sense, we have named the relation~(\ref{eq:sol-13}) as the "Solar-CHOOZ Complementarity". However, in contrast to the Quark-Lepton Complementarity~\cite{Minakata:2004xt}, Eq.~(\ref{eq:sol-13}) indicates that only if the CHOOZ angle $\theta_{13}$ is sizable, the solar angle $\theta_{sol}$ can deviate from the maximal mixing $\pi/4$. For $\lambda = 0.22$~\cite{Eidelman:2004wy}, our predictions are 
\begin{eqnarray}
 \theta_{atm} = 45^\circ + {\cal O}(\lambda^4), \quad
 \theta_{sol} = 36^\circ + {\cal O}(\lambda^4), \quad
 \theta_{13} = 9^\circ   + {\cal O}(\lambda^3),
\end{eqnarray}
which are consistent with experiments~\cite{Maltoni:2004ei}, and $\lambda^4 (\lambda^3)$ radians $\simeq 0.1^\circ (0.6^\circ)$ for $\lambda = 0.22$~\cite{Eidelman:2004wy}. Furthermore, Eq.~(\ref{eq:sol-13}) and the experimental upper bound on the solar angle $\theta_{sol}$~\cite{Maltoni:2004ei} give the lower bound on the CHOOZ angle $\theta_{13}$
\begin{eqnarray}
 \sin^2 \theta_{sol} < 0.38 \ (0.41) \quad \longrightarrow \quad 7^\circ \ (5^\circ) < \theta_{13},
\end{eqnarray}
where $3 \sigma \ (4 \sigma)$ for the experimental bound~\cite{Maltoni:2004ei}.

We also infer the CP violation in neutrino oscillations. In our model, from Eq.~(\ref{eq:MNS2}), the standard Jarlskog CP violation $J$ factor~\cite{Jarlskog:1985ht} is given by
\begin{eqnarray}
 J_{MNS} \simeq \frac{1}{4\sqrt{2}} A \lambda^3 \eta,   \label{eq:JMNS1}
\end{eqnarray}
where $A$, $\lambda$ and $\eta$ are CKM quark mixing parameters in Wolfenstein parametrization~\cite{Wolfenstein:1983yz}. We can also write the Jarlskog CP violation $J$ factor only with the quantities in Eq.~(\ref{eq:angles})~\cite{Jarlskog:1985ht}
\begin{eqnarray}
 J_{MNS} = \frac{1}{4}\sqrt{\sin^2 2\theta_{sol}} \sqrt{\sin^2 2\theta_{atm}} \sqrt{\sin^2 \theta_{13}}(1-\sin^2 \theta_{13}) \sin \delta_{MNS},   \label{eq:JMNS2}
\end{eqnarray}
where $\delta_{MNS}$ is the leptonic Dirac CP phase.
Therefore, from Eqs.~(\ref{eq:angles}),~(\ref{eq:JMNS1}) and~(\ref{eq:JMNS2}), the leptonic Dirac CP phase is predicted as 
\begin{eqnarray}
 \sin \delta_{MNS} \simeq A \lambda^2 \eta.  \label{eq:phase}
\end{eqnarray}
Eq.~(\ref{eq:phase}) indicates that in contrast to the quark CP phase $\delta_{CKM} = 60^\circ \pm 14^\circ$~\cite{Eidelman:2004wy}, the leptonic Dirac CP phase is very small,
\begin{eqnarray}
 \delta_{MNS} \simeq 0.8^\circ,
\end{eqnarray}
where we have used $\lambda = 0.22$, $A = 0.85$ and $\overline{\eta} = 0.33$ with $\overline{\eta} \equiv \eta (1-\lambda^2/2)$~\cite{Eidelman:2004wy}.

We finally remark that the ratio of the Jarlskog CP violation $J$ factor for quarks and leptons $R_J \equiv J_{CKM}/J_{MNS}$ is important, because the large uncertainty on $\eta$ is cancelled out in the ratio $R_J$,
\begin{eqnarray}
 R_J \equiv \frac{J_{CKM}}{J_{MNS}} \simeq 4 \sqrt{2} A \lambda^3 \simeq 5 \times 10^{-2}, \label{eq:Ratio}
\end{eqnarray}
where $J_{CKM} \simeq A^2 \lambda^6 \eta$~\cite{Jarlskog:1985ht}. Thus, the ratio $R_J$ does not depend on the CP violation parameter $\eta$ and is determined only by the mixing angle parameters $A$ and $\lambda$ with less uncertainty. Although $\delta_{CKM} \gg \delta_{MNS}$, $R_J \ll 1$ is satisfied due to the smallness of the CKM quark flavour mixing angles.

In conclusion, we have demonstrated that only two ansatz~(\ref{eq:1ansatz}) and~(\ref{eq:2ansatz}) are sufficient to produce the neutrino mixing angles, and presented their analytical expressions~(\ref{eq:angles}). We have also discussed the CP violation in neutrino oscillations. If our predictions of the CP violation~(\ref{eq:Ratio}) and the "Solar-CHOOZ Complementarity" relation~(\ref{eq:sol-13}) will be confirmed by the future experiments, we should take our two ansatz~(\ref{eq:1ansatz}) and~(\ref{eq:2ansatz}) seriously and investigate the theoretical origin of the second ansatz~(\ref{eq:2ansatz}).

\acknowledgements
The author thanks the Asia Pacific Center for Theoretical Physics for financial support.

\end{document}